\documentclass[lettersize,journal]{IEEEtran}
\usepackage{amsmath,amsfonts}
\usepackage{algorithmic}
\usepackage{algorithm}
\usepackage{array}
\usepackage[caption=false,font=normalsize,labelfont=sf,textfont=sf]{subfig}
\usepackage{textcomp}
\usepackage{stfloats}
\usepackage{url}
\usepackage{verbatim}
\usepackage{graphicx}
\usepackage{cite}
\usepackage{multirow}
\begin{document}

\title{WiFo-D: Wireless Foundation Model Based on Discrete Representations }

\author{Xinyu Pan,~\IEEEmembership{Graduate Student Member,~IEEE,}
        Boxun Liu,~\IEEEmembership{Graduate Student Member,~IEEE,}
        Chen Chen,~\IEEEmembership{Senior Member,~IEEE,}
        and Xiang Cheng,~\IEEEmembership{Fellow,~IEEE}
\thanks{Xinyu Pan, Boxun Liu, Chen Chen, and Xiang Cheng are with the
State Key Laboratory of Photonics and Communications, School of
Electronics, Peking University, Beijing 100871, China
(e-mail: 2501213461@stu.pku.edu.cn; boxunliu@stu.pku.edu.cn;
c.chen@pku.edu.cn; xiangcheng@pku.edu.cn).}
}

\markboth{WiFo-D: Research Preprint}%
{Pan \MakeLowercase{\textit{et al.}}: WiFo-D: Wireless Foundation Model Based on Discrete Representations}

\maketitle

\begin{abstract}
Wireless foundation models have shown strong generalization capabilities across heterogeneous wireless tasks and configurations. However, most existing models rely on continuous representations, limiting ultra-low-overhead channel state information (CSI) feedback, few-shot generalization, and robustness to noise. To address these limitations, this paper proposes a wireless foundation model based on discrete representations (WiFo-D), which learns multi-granularity discrete CSI representations. It comprises a CSI encoder, multi-codebook vector quantization, and a modulation-based decoder, with the learned token maps and quantized features supporting multi-rate CSI feedback, CSI reconstruction, and diverse CSI-related downstream tasks. Extensive experiments demonstrate strong CSI feedback performance at an ultra-low latent compression ratio of 1/32 with 3 bits per token, reducing normalized mean square error (NMSE) by more than 4.7 dB compared with the strongest baseline. Furthermore, for frequency-domain prediction with limited training samples, WiFo-D improves generalization and noise robustness over continuous-representation-based methods. The results also demonstrate its superiority across downstream tasks, including beam prediction, wireless localization, and line-of-sight/non-line-of-sight identification.
\end{abstract}

\begin{IEEEkeywords}
Wireless foundation model, channel state information (CSI), vector
quantization, CSI feedback, CSI reconstruction.
\end{IEEEkeywords}

\section{Introduction}
\IEEEPARstart{F}{uture} wireless networks are evolving from merely connecting people and devices toward supporting ubiquitous intelligent services, such as immersive communications, digital twins, intelligent transportation, and embodied intelligence, which impose stringent requirements on physical-layer modeling under complex propagation environments, diverse terminal devices, and heterogeneous system configurations~\cite{ref_6G_Vision, ref_SoM, ref_EIW}. To mitigate the model mismatch suffered by traditional rule-based or model-driven methods, learning-based techniques have been widely explored and have achieved promising performance in various physical-layer tasks, including channel estimation~\cite{ref_Channelformer}, channel state information (CSI) feedback~\cite{ref_CsiNet}, channel prediction~\cite{ref_Transformer_CP}, and beam prediction~\cite{ref_DL_BP}, among others. However, most existing learning-based wireless models are still designed for specific tasks or system configurations, making their learned knowledge difficult to reuse across heterogeneous scenarios. Once the propagation environment or task objective changes, dedicated retraining or fine-tuning is usually required, leading to fragmented model designs and limited transferability.

Inspired by the generalization, transfer, and task adaptation
capabilities of foundation models in natural language processing,
computer vision, and multimodal learning~\cite{ref_GPT3, ref_MAE, ref_CLIP},
recent studies have explored native wireless foundation models
trained on large-scale wireless data~\cite{ref_LWFM, ref_FM_SoM}.
Early works explored complementary adaptation capabilities. LWM focused
on data-efficient adaptation under limited training data, while the
contemporaneous WiFo (Wireless Foundation Model) first demonstrated zero-shot channel prediction
across heterogeneous configurations~\cite{ref_LWM,ref_WiFo}.
Building on WiFo, subsequent works in the WiFo series extended such
generalization to broader wireless tasks, including heterogeneous
CSI feedback with WiFo-CF~\cite{ref_WiFo_CF}, end-to-end FDD
precoding with WiFo-E~\cite{ref_WiFo_E}, and unified communication
and sensing across scenarios, configurations, and tasks with
WiFo-2~\cite{ref_WiFo2}. The WiFo series has also been extended to
multimodal sensing-communication integration~\cite{ref_WiFo_M2,ref_WiFo_MiSAC}. Other studies have explored
generative multi-task learning, real-world channel foundation
modeling, transferable channel representations, in-context
physical-layer adaptation, and environment-aware multi-task modeling~\cite{ref_WirelessGPT,ref_ChannelGPT,ref_CSI_CLIPpp,
ref_ICWLM,ref_MUSE_FM}.
These studies indicate that wireless foundation models provide a
promising paradigm for improving knowledge reuse, task adaptation,
and generalization across diverse wireless tasks, configurations,
and scenarios. From a representation perspective, most existing
wireless foundation models learn continuous latent spaces, where CSI
or wireless observations are encoded into continuous embeddings or
feature maps. Such continuous representations are effective in
capturing fine-grained amplitude, phase, and space-time-frequency
correlations of wireless channels, and have shown strong performance
in various CSI-related tasks.

However, continuous-representation-based wireless foundation
models still face two inherent representation-level
limitations. First, continuous latent features remain
high-dimensional and are not inherently compact for finite-bit
interfaces. For CSI compression feedback, they must therefore
be converted into discrete messages through an additional
quantization stage~\cite{ref_WiFo_CF}. Under stringent
feedback budgets, directly quantizing an unconstrained
continuous space may discard correlated antenna-frequency
structures and degrade CSI reconstruction accuracy. Second, continuous feature spaces lack explicit structural
priors that organize CSI into representative channel patterns.
Models therefore learn and infer directly in an unconstrained
continuous space, which may require more task-specific samples
to learn the CSI reconstruction mapping and may be more susceptible
to perturbations, distribution shifts, and noisy observations.
For CSI reconstruction, these properties may manifest as reduced
sample efficiency, weaker cross-domain transfer, and lower
robustness to corrupted observations. For downstream inference,
reconstructed CSI or continuous CSI features may also introduce
representation mismatch and require additional task-model
adaptation~\cite{ref_WiFo2,ref_WiFo_CF}. In FDD systems, where
complete downlink CSI is unavailable at the base station and
must be acquired through feedback, estimation, or prediction,
the lack of a compact and structured representation interface
may further introduce distribution mismatch between CSI
reconstruction and downstream inference, require additional
task-model adaptation, and increase deployment complexity.

These limitations suggest that the representation form itself should be reconsidered. In computer vision and multimodal learning, discrete representations learned by vector-quantized codebooks have demonstrated the feasibility of transforming continuous signals into compact tokens, enabling efficient compression, masked modeling, and generative completion~\cite{ref_VQVAE, ref_VQGAN, ref_BEiT}. In wireless communications, vector quantization (VQ) and vector-quantized variational autoencoder (VQ-VAE)-based methods have also been explored for CSI-related feedback, including finite-bit CSI feedback, channel-adaptive feedback, and precoding-oriented feedback design~\cite{ref_VQ_CSI_Feedback, ref_EC_VQVAE_CSI, ref_VQ_DJSCC_CSI, ref_VQ_Precoding_MI, ref_VQVAE_Robust_Precoding}. These studies validate the effectiveness of discrete codebooks in wireless systems. However, they are usually developed from specific task objectives, such as CSI compression, channel reconstruction, or precoding performance optimization, rather than from the perspective of learning general-purpose discrete CSI representations. As a result, the learned codebooks are often tied to particular feedback objectives, system settings, or deployment conditions, which limits their potential to serve as a general representation interface across heterogeneous wireless tasks and configurations. For wireless foundation models, discrete CSI representations are expected to play a more fundamental role, since CSI samples with similar propagation structures can be mapped to shared codewords, allowing the codebook to capture typical channel patterns as compact structural priors. The resulting discrete indices are naturally compatible with finite-bit feedback, while token-level modeling can constrain CSI reconstruction within a learned set of representative channel patterns instead of directly regressing in a high-dimensional continuous space. Despite this potential, how to learn general-purpose discrete CSI codebooks from large-scale heterogeneous wireless data and how to support multi-rate CSI feedback, discrete-token-based CSI reconstruction, and downstream task adaptation with such tokens remain largely unexplored.

To address this problem, this paper proposes a wireless
foundation model based on discrete representations, termed
WiFo-D. WiFo-D aims to learn a general-purpose discrete CSI
token space from large-scale heterogeneous wireless data,
rather than a task-specific codebook tied to a particular
feedback objective, system configuration, or deployment
scenario. The discrete representation-learning component of WiFo-D
consists of a CSI encoder, a multi-codebook vector
quantization module, and a modulation-based CSI decoder. The encoder extracts latent
channel features, while codebooks with different sizes produce
discrete token maps and quantized feature maps at multiple
granularities. The decoder reconstructs CSI from the selected
quantized representation under a shared backbone, thereby
supporting multi-rate CSI feedback. Beyond CSI feedback, WiFo-D supports CSI reconstruction from partial observations
by generating a complete CSI token map conditioned on the
available observations. In this work,
we instantiate this general formulation with frequency-domain
CSI prediction as a representative reconstruction task. A
lightweight predictor takes the quantized feature map extracted
from partially observed CSI and generates a complete token map,
which is subsequently converted into a quantized feature map
and decoded into reconstructed CSI. The
learned quantized features can also be directly reused as compact
representations for downstream wireless tasks.

The main contributions of this paper are summarized as follows.

\begin{itemize}
\item We propose WiFo-D, a wireless foundation model that
constructs a multi-granularity discrete CSI representation
space from heterogeneous CSI data. A multi-codebook vector
quantization module and a modulation-based decoder jointly
capture representative channel patterns at different
granularities under a shared backbone.

\item We develop task-adaptation mechanisms based on the
learned discrete representations. Codebooks with different
sizes enable native multi-rate CSI feedback. A lightweight
predictor generates complete token maps from observed
quantized features for CSI reconstruction, instantiated here by
frequency-domain prediction. The quantized features can also
be directly fed into lightweight downstream task heads.

\item Extensive experiments demonstrate strong low-bit in-distribution
(ID) and out-of-distribution (OOD) CSI feedback performance
and flexible 3-bit to 8-bit operation. For frequency-domain
prediction, WiFo-D improves sample efficiency, representative
cross-domain transfer, and noise robustness, while its quantized
features achieve superior performance in beam prediction,
wireless localization, and LoS/NLoS identification. Ablation,
scaling, and complexity analyses further validate the key designs
and deployment efficiency of WiFo-D.
\end{itemize}

\emph{Notation:} Scalars, vectors, matrices, and sets are denoted
by italic, bold lowercase, bold uppercase, and calligraphic
symbols, respectively. The real and complex fields are denoted by
$\mathbb{R}$ and $\mathbb{C}$. The operators
$\Re\{\cdot\}$ and $\Im\{\cdot\}$ denote the real and
imaginary parts, respectively. The entry-wise $\ell_1$ norm, Euclidean norm, and Frobenius norm are denoted by
$\lVert\cdot\rVert_1$, $\lVert\cdot\rVert_2$, and
$\lVert\cdot\rVert_F$, respectively. The cardinality of a set is
denoted by $|\cdot|$, and $\mathbb{E}\{\cdot\}$ denotes
expectation.

\section{System Model and Problem Formulation}
This section defines the heterogeneous multiple-input single-output orthogonal frequency
division multiplexing (MISO-OFDM) CSI data
model and formulates the discrete representation interfaces for
CSI feedback, frequency-domain prediction, and downstream task
adaptation.

\subsection{CSI Data Model}

We consider a MISO-OFDM system in which the base
station (BS) employs an $N_{\rm h}\times N_{\rm v}$ uniform
planar array (UPA), while the user equipment (UE) is equipped
with a single antenna. The total number of BS antennas is
$N_t=N_{\rm h}N_{\rm v}$. The frequency-domain channel vector
$\mathbf{h}(f)\in\mathbb{C}^{N_t}$ between the BS and UE at
frequency $f$ is modeled as
\begin{equation}
\mathbf{h}(f)
=
\sum_{m=1}^{N_{\rm cl}}
\sum_{p=1}^{P_m}
\beta_{m,p}
e^{-j2\pi f\tau_{m,p}}
e^{j\Phi_{m,p}}
\boldsymbol{\alpha}
\left(
\theta_{m,p},
\phi_{m,p}
\right),
\end{equation}
where $N_{\rm cl}$ and $P_m$ denote the number of clusters
and the number of paths in the $m$-th cluster, respectively.
Moreover, $\beta_{m,p}$, $\tau_{m,p}$, and $\Phi_{m,p}$
denote the path amplitude, delay, and random phase of the
corresponding path, respectively, while $\theta_{m,p}$ and
$\phi_{m,p}$ denote the azimuth and elevation angles. The UPA
steering vector satisfies
$\boldsymbol{\alpha}(\theta_{m,p},\phi_{m,p})
\in\mathbb{C}^{N_t}$.

Let $K$ denote the number of uniformly spaced frequency
samples and let $\Delta f$ denote the frequency sampling
interval. Assuming that the frequency samples are symmetric
about the carrier frequency $f_c$, the first sampled frequency
is
\begin{equation}
f_1
=
f_c-\frac{K-1}{2}\Delta f.
\end{equation}
The channel vector at the $k$-th frequency sample and the
resulting CSI matrix are given by
\begin{equation}
\begin{gathered}
\mathbf{h}_k
=
\mathbf{h}
\left(
f_1+(k-1)\Delta f
\right)
\in
\mathbb{C}^{N_t},
\quad k=1,\ldots,K,\\
\mathbf{H}
=
\left[
\mathbf{h}_1,
\mathbf{h}_2,
\ldots,
\mathbf{h}_K
\right]
\in
\mathbb{C}^{N_t\times K}.
\end{gathered}
\end{equation}
The corresponding nominal bandwidth is $B=K\Delta f$.

For heterogeneous pre-training, we consider the CSI dataset
collection
\begin{equation}
\mathcal{D}
=
\left\{
\mathcal{D}_d
\right\}_{d=1}^{N_{\rm d}},
\qquad
\mathbf{H}_i^{(d)}
\in
\mathbb{C}^{
N_t^{(d)}
\times
K^{(d)}
},
\end{equation}
where $N_{\rm d}$ is the number of sub-datasets,
$\mathbf{H}_i^{(d)}$ is the $i$-th sample in
$\mathcal{D}_d$, and
$N_t^{(d)}=N_{\rm h}^{(d)}N_{\rm v}^{(d)}$ and $K^{(d)}$
denote the numbers of BS antennas and frequency samples under
the $d$-th configuration, respectively.
The sub-datasets differ in propagation scenario, carrier
frequency, bandwidth, frequency spacing, antenna configuration,
and UE distribution, while sharing common multipath channel
structures.

\subsection{Problem Formulation}

Based on the heterogeneous CSI data model above, WiFo-D aims
to learn a shared multi-granularity discrete representation
space that can be reused across different wireless tasks and
system configurations. A CSI sample is first mapped to a latent
feature map and then quantized into a discrete token map and its
corresponding quantized feature map. The token map provides a
finite-index interface for CSI feedback and token prediction,
whereas the quantized feature map can be decoded into CSI or
directly reused for downstream inference. Based on this shared
representation, the CSI representation and task interfaces are
formulated as follows.

\subsubsection{Unified Discrete CSI Representation}

Given a CSI sample $\mathbf{H}$, let $F_{\rm e}(\cdot)$ denote
the CSI representation mapping, and let
$Q^{(r)}(\cdot;\mathcal{C}^{(r)})$ denote the vector
quantization mapping associated with the $r$-th codebook.
The latent feature map and the corresponding discrete
representations are obtained as
\begin{equation}
\begin{gathered}
\mathbf{Z}
=
F_{\rm e}
\left(
\mathbf{H}
\right)
\in
\mathbb{R}^{d_{\rm c}\times H_z\times W_z},\\
\left(
\mathbf{S}^{(r)},
\mathbf{Z}_{\rm q}^{(r)}
\right)
=
Q^{(r)}
\left(
\mathbf{Z};
\mathcal{C}^{(r)}
\right),
\qquad
r=1,\ldots,R,
\end{gathered}
\end{equation}
where $R$ is the number of codebooks,
$\mathcal{C}^{(r)}$ contains
$N_{\rm c}^{(r)}=|\mathcal{C}^{(r)}|$ codewords of dimension
$d_{\rm c}$,
$\mathbf{S}^{(r)}
\in\{1,\ldots,N_{\rm c}^{(r)}\}^{H_z\times W_z}$ is the
discrete token map, and
$\mathbf{Z}_{\rm q}^{(r)}
\in\mathbb{R}^{d_{\rm c}\times H_z\times W_z}$ is the
corresponding quantized feature map. Here, $H_z$ and $W_z$
denote the latent spatial dimensions. Codebooks with different
sizes provide discrete CSI representations at different
granularities under the same representation mapping.

\subsubsection{Task Interfaces}

Based on the shared discrete representation, WiFo-D supports
CSI compression feedback, frequency-domain CSI prediction, and
CSI-related downstream task adaptation.

For CSI compression feedback, let $L=H_zW_z$ denote the
number of token positions, and let $F_{\rm d}(\cdot)$ denote
the CSI reconstruction mapping from a quantized feature map to
complex-valued CSI. The feedback overhead and reconstructed CSI
under the $r$-th codebook are expressed as
\begin{equation}
\begin{gathered}
N_{\rm bit}^{(r)}
=
L
\left\lceil
\log_2 N_{\rm c}^{(r)}
\right\rceil,\\
\hat{\mathbf{H}}^{(r)}
=
F_{\rm d}
\left(
\mathbf{Z}_{\rm q}^{(r)}
\right).
\end{gathered}
\end{equation}
Only the discrete token map $\mathbf{S}^{(r)}$ is transmitted,
while the BS recovers $\mathbf{Z}_{\rm q}^{(r)}$ through
codebook lookup. Selecting codebooks with different sizes
therefore provides different feedback rates under the same
encoder-decoder backbone.

For CSI reconstruction from partial frequency observations, let
$\Omega\subseteq\{1,\ldots,K\}$ denote the observed frequency
indices and let $\bar{\Omega}$ denote their complement. The
partially observed CSI $\mathbf{H}_{\Omega}$ retains the
complete antenna-domain channel vectors
$\mathbf{H}(:,k)$ for $k\in\Omega$, while the entire frequency
columns indexed by $k\in\bar{\Omega}$ are masked. Thus, the
masking operation is applied only along the frequency
dimension, whereas all $N_t$ antenna coefficients at each
observed frequency remain available. The quantized feature map
$\mathbf{Z}_{{\rm q},\Omega}^{(r)}$ is then extracted from
$\mathbf{H}_{\Omega}$ using the shared representation and
tokenization mappings. A task-specific predictor
$G_{\rm f}^{(r)}(\cdot)$ subsequently generates codeword
probabilities over all positions of the complete token map:
\begin{equation}
\mathbf{P}^{(r)}
=
G_{\rm f}^{(r)}
\left(
\mathbf{Z}_{{\rm q},\Omega}^{(r)}
\right)
\in
[0,1]^{L\times N_{\rm c}^{(r)}}.
\end{equation}
The predicted complete token map and reconstructed CSI are
obtained as
\begin{equation}
\begin{gathered}
\widetilde{\mathbf{S}}^{(r)}(i)
=
\arg\max_{1\leq j\leq N_{\rm c}^{(r)}}
\left[
\mathbf{p}_i^{(r)}
\right]_j,
\qquad i\in\mathcal{I},\\
\widetilde{\mathbf{H}}^{(r)}
=
F_{\rm d}
\left(
\operatorname{Lookup}
\left(
\widetilde{\mathbf{S}}^{(r)};
\mathcal{C}^{(r)}
\right)
\right),
\end{gathered}
\end{equation}
where
$\mathbf{p}_i^{(r)}\in[0,1]^{N_{\rm c}^{(r)}}$ is the
$i$-th row of $\mathbf{P}^{(r)}$ and represents the predicted
codeword distribution at token position $i$,
$\mathcal{I}=\{1,\ldots,L\}$ denotes all positions in the
complete token map, and
$\operatorname{Lookup}(\mathbf{S};\mathcal{C})$ retrieves
from codebook $\mathcal{C}$ the codeword indexed by each token
in $\mathbf{S}$ and arranges the retrieved codewords into the
corresponding quantized feature map.

For downstream adaptation, the quantized feature map serves as
a compact task representation and is directly fed into a
lightweight task head without first reconstructing CSI:
\begin{equation}
\hat{\mathbf{y}}
=
h_i
\left(
\mathbf{Z}_{\rm q}^{(r)}
\right),
\qquad
i\in\{1,2,3\},
\end{equation}
where $h_1$, $h_2$, and $h_3$ correspond to beam prediction,
wireless localization, and LoS/NLoS identification,
respectively.

The representation-learning architecture and the corresponding
task-adaptation mechanisms are detailed in Sections III and IV,
respectively.

\section{Discrete CSI Representation Learning of WiFo-D}

\begin{figure*}[t]
\centering
\includegraphics[width=0.92\textwidth]{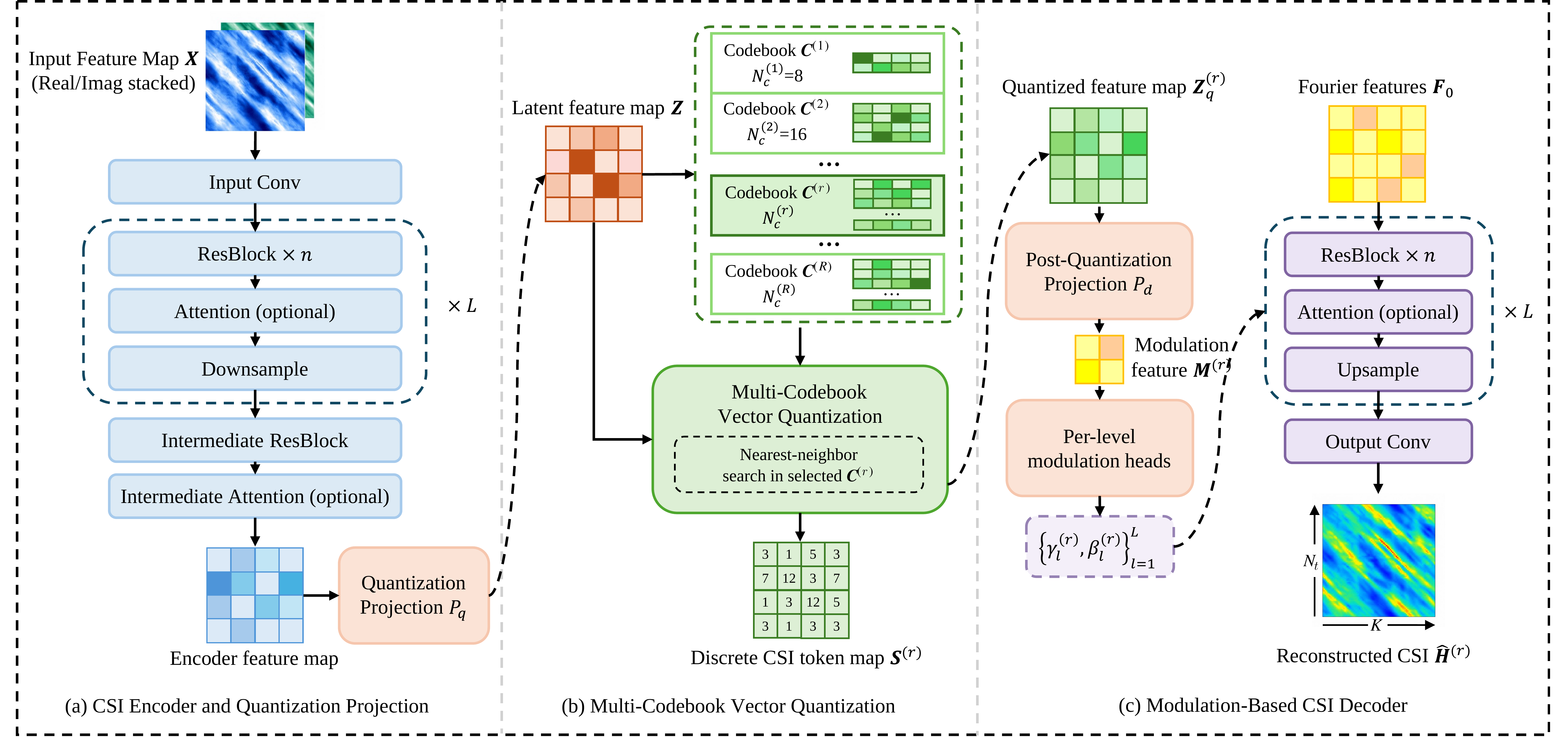}
\caption{Architecture for discrete CSI representation learning in WiFo-D.}
\label{fig:wifo_d_details}
\end{figure*}

This section presents how WiFo-D learns multi-granularity
discrete CSI representations. As illustrated in
Fig.~\ref{fig:wifo_d_details}, the architecture for discrete
CSI representation learning consists of three main modules: the CSI encoder
and quantization projection, multi-codebook vector quantization,
and modulation-based CSI decoder. These modules implement the
mappings $F_{\rm e}(\cdot)$, $Q^{(r)}(\cdot)$, and
$F_{\rm d}(\cdot)$, respectively, and are trained under a
multi-codebook self-supervised pre-training strategy. Multiple codebooks with different sizes share the same encoder,
projection modules, and decoder, enabling WiFo-D to learn
multi-granularity discrete CSI representations for task
adaptation in Section~IV.

\subsection{CSI Encoder and Quantization Projection}

As illustrated in Fig.~\ref{fig:wifo_d_details}(a), the CSI
encoder and quantization projection jointly implement the
representation mapping $F_{\rm e}(\cdot)$. The encoder extracts
features over the antenna-frequency grid, while the quantization
projection maps the encoder output to the codeword dimension.

Given
$\mathbf{H}\in\mathbb{C}^{N_t\times K}$, its real and imaginary
parts are stacked as
\begin{equation}
\mathbf{X}
=
\mathcal{T}(\mathbf{H})
=
\operatorname{stack}
\left(
\Re{\mathbf{H}},
\Im{\mathbf{H}}
\right)
\in
\mathbb{R}^{2\times N_t\times K}.
\end{equation}
Here, $\mathcal{T}(\cdot)$ denotes real-imaginary stacking,
while $\mathcal{T}^{-1}(\cdot)$ denotes the inverse mapping
from the two-channel real representation to complex-valued
CSI.

The encoder $E_{\theta}(\cdot)$ and quantization projection
$P_{\rm q}(\cdot)$ produce
\begin{equation}
\mathbf{Z}
=
F_{\rm e}(\mathbf{H})
=
P_{\rm q}
\left(
E_{\theta}
\left(
\mathbf{X}
\right)
\right)
\in
\mathbb{R}^{d_{\rm c}\times H_z\times W_z},
\end{equation}
where $d_{\rm c}$ is the codeword dimension, and $H_z$ and
$W_z$ are the latent spatial dimensions. The encoder contains
convolutional layers, residual blocks, downsampling modules,
and attention blocks to capture local and global dependencies
over the antenna-frequency grid.

\subsection{Multi-Codebook Vector Quantization}

As illustrated in Fig.~\ref{fig:wifo_d_details}(b), the
multi-codebook vector quantization module
$Q^{(r)}(\cdot)$ assigns each latent vector to its nearest
codeword in the selected codebook. The resulting codeword
indices form the discrete token map $\mathbf{S}^{(r)}$, while
the corresponding codewords form the quantized feature map
$\mathbf{Z}_{\rm q}^{(r)}$.

WiFo-D maintains $R$ independent codebooks,
\begin{equation}
\mathcal{C}^{(r)}
=
\left\{
\mathbf{e}_j^{(r)}
\right\}_{j=1}^{N_{\rm c}^{(r)}},
\qquad
\mathbf{e}_j^{(r)}
\in
\mathbb{R}^{d_{\rm c}},
\end{equation}
where $N_{\rm c}^{(r)}$ is the number of codewords in the
$r$-th codebook. For each latent vector
$\mathbf{z}_{u,v}\in\mathbb{R}^{d_{\rm c}}$, the token index
is determined by nearest-neighbor assignment:
\begin{equation}
s_{u,v}^{(r)}
=
\arg\min_{1\leq j\leq N_{\rm c}^{(r)}}
\left\|
\mathbf{z}_{u,v}
-
\mathbf{e}_j^{(r)}
\right\|_2^2.
\end{equation}
The discrete token and its quantized feature at position
$(u,v)$ are then given by
\begin{equation}
\begin{gathered}
\mathbf{S}^{(r)}(u,v)
=
s_{u,v}^{(r)},\\[2pt]
\mathbf{z}_{{\rm q},u,v}^{(r)}
=
\mathbf{e}_{s_{u,v}^{(r)}}^{(r)}.
\end{gathered}
\end{equation}
Collecting all positions yields
$\left(\mathbf{S}^{(r)},\mathbf{Z}_{\rm q}^{(r)}\right)
=Q^{(r)}(\mathbf{Z};\mathcal{C}^{(r)})$, where
$\mathbf{Z}_{\rm q}^{(r)}
\in\mathbb{R}^{d_{\rm c}\times H_z\times W_z}$.

\subsection{Modulation-Based CSI Decoder}

As illustrated in Fig.~\ref{fig:wifo_d_details}(c), the CSI
reconstruction mapping $F_{\rm d}(\cdot)$ is implemented by a
modulation-based decoder inspired by
MoVQ~\cite{ref_MoVQ}. Instead of using the quantized feature
map only as the initial decoder activation,
$\mathbf{Z}_{\rm q}^{(r)}$ modulates intermediate decoder
features through conditional normalization. The decoder
therefore takes an initial feature $\mathbf{F}_0$ and a
sample-dependent quantized conditioning feature as two
complementary inputs.

Following MoVQ, we consider positional features, learned
constants, and Fourier features as candidate initializations of
$\mathbf{F}_0$~\cite{ref_MoVQ}. We additionally consider
transformed quantized features following
Emu3~\cite{ref_Emu3}. WiFo-D adopts Fourier features, which
provide a coordinate-dependent basis for the antenna-frequency
grid, while the quantized feature map carries sample-dependent
channel information.

The quantized feature map is first transformed by the
post-quantization projection and resized for each decoder block:
\begin{equation}
\begin{gathered}
\mathbf{M}^{(r)}
=
P_{\rm d}
\left(
\mathbf{Z}_{\rm q}^{(r)}
\right),\\
\mathbf{M}_{\ell}^{(r)}
=
\operatorname{Resize}_{\ell}
\left(
\mathbf{M}^{(r)}
\right),
\end{gathered}
\end{equation}
where $P_{\rm d}(\cdot)$ denotes the post-quantization
projection, and $\operatorname{Resize}_{\ell}(\cdot)$ matches
the spatial resolution of the $\ell$-th decoder block.

Let $\mathbf{F}_{\ell}$ denote the input activation of the
$\ell$-th decoder block. The resized conditioning feature
generates the block-wise scaling and shifting parameters:
\begin{equation}
\begin{gathered}
\boldsymbol{\gamma}_{\ell}^{(r)}
=
G_{\ell}^{\gamma}
\left(
\mathbf{M}_{\ell}^{(r)}
\right),\\
\boldsymbol{\beta}_{\ell}^{(r)}
=
G_{\ell}^{\beta}
\left(
\mathbf{M}_{\ell}^{(r)}
\right),
\end{gathered}
\end{equation}
where $G_{\ell}^{\gamma}(\cdot)$ and
$G_{\ell}^{\beta}(\cdot)$ are lightweight projection
functions. The normalized activation is modulated as
\begin{equation}
\widetilde{\mathbf{F}}_{\ell}
=
\boldsymbol{\gamma}_{\ell}^{(r)}
\odot
\operatorname{Norm}
\left(
\mathbf{F}_{\ell}
\right)
+
\boldsymbol{\beta}_{\ell}^{(r)},
\end{equation}
where $\operatorname{Norm}(\cdot)$ denotes normalization and
$\odot$ denotes element-wise multiplication.
The modulated features are processed by residual, attention, and
upsampling blocks to recover the original antenna-frequency
resolution. The decoder output and complex-valued CSI
reconstruction are
\begin{equation}
\begin{gathered}
\hat{\mathbf{X}}^{(r)}
=
D_{\psi}
\left(
\mathbf{F}_0;
\mathbf{Z}_{\rm q}^{(r)}
\right)
\in
\mathbb{R}^{2\times N_t\times K},\\
\hat{\mathbf{H}}^{(r)}
=
F_{\rm d}
\left(
\mathbf{Z}_{\rm q}^{(r)}
\right)
=
\mathcal{T}^{-1}
\left(
\hat{\mathbf{X}}^{(r)}
\right).
\end{gathered}
\end{equation}
Here, $D_{\psi}(\cdot;\cdot)$ denotes the real-valued decoder
network parameterized by $\psi$, with $\mathbf{F}_0$ serving
as the initial decoder feature and
$\mathbf{Z}_{\rm q}^{(r)}$ providing sample-dependent
conditioning.

\subsection{Multi-Codebook Self-Supervised Pre-Training}

WiFo-D learns discrete CSI representations from the
heterogeneous dataset collection $\mathcal{D}$ using CSI
reconstruction as self-supervision. The encoder, projection
modules, and decoder are shared by all codebooks, while the
$R$ codebooks are independent embedding tables with different
sizes. During each iteration, one active codebook is sampled
according to
\begin{equation}
r
\sim
p_{\rm cb}(r),
\qquad
r\in\{1,\ldots,R\},
\end{equation}
where $p_{\rm cb}(r)$ is uniform in the default setting. Only the selected codebook participates in the current
quantization operation, while the shared backbone is optimized
across different codebook granularities throughout training.

Since nearest-neighbor assignment is non-differentiable,
WiFo-D follows VQ-VAE~\cite{ref_VQVAE} and applies the
straight-through estimator. Let $\operatorname{sg}(\cdot)$
denote the stop-gradient operator. For the active codebook, the reconstruction loss is defined as
\begin{equation}
\ell_{\rm rec}^{(r)}
=
\frac{1}{2N_tK}
\left\|
\mathbf{X}
-
\hat{\mathbf{X}}^{(r)}
\right\|_1.
\end{equation}

The codebook and commitment losses are respectively given by
\begin{equation}
\begin{gathered}
\ell_{\rm cb}^{(r)}
=
\frac{1}{d_{\rm c}H_zW_z}
\left\|
\operatorname{sg}(\mathbf{Z})
-
\mathbf{Z}_{\rm q}^{(r)}
\right\|_F^2,\\[3pt]
\ell_{\rm com}^{(r)}
=
\frac{1}{d_{\rm c}H_zW_z}
\left\|
\mathbf{Z}
-
\operatorname{sg}
\left(
\mathbf{Z}_{\rm q}^{(r)}
\right)
\right\|_F^2.
\end{gathered}
\end{equation}

The overall pre-training objective is
\begin{equation}
\ell_{\rm pre}^{(r)}
=
\ell_{\rm rec}^{(r)}
+
\lambda_{\rm cb}\ell_{\rm cb}^{(r)}
+
\lambda_{\rm com}\ell_{\rm com}^{(r)},
\end{equation}
where $\lambda_{\rm cb}$ and $\lambda_{\rm com}$ control the
weights of the codebook and commitment losses, respectively.
The reconstruction loss trains the shared encoder, projection
modules, and decoder, the codebook loss updates the active
codewords, and the commitment loss constrains encoder features
to remain close to their assigned codewords. Alternating among codebooks
therefore produces multi-granularity discrete CSI
representations under one shared backbone.

\section{Task Adaptation of WiFo-D}

\begin{figure*}[t]
\centering
\includegraphics[width=0.90\textwidth]{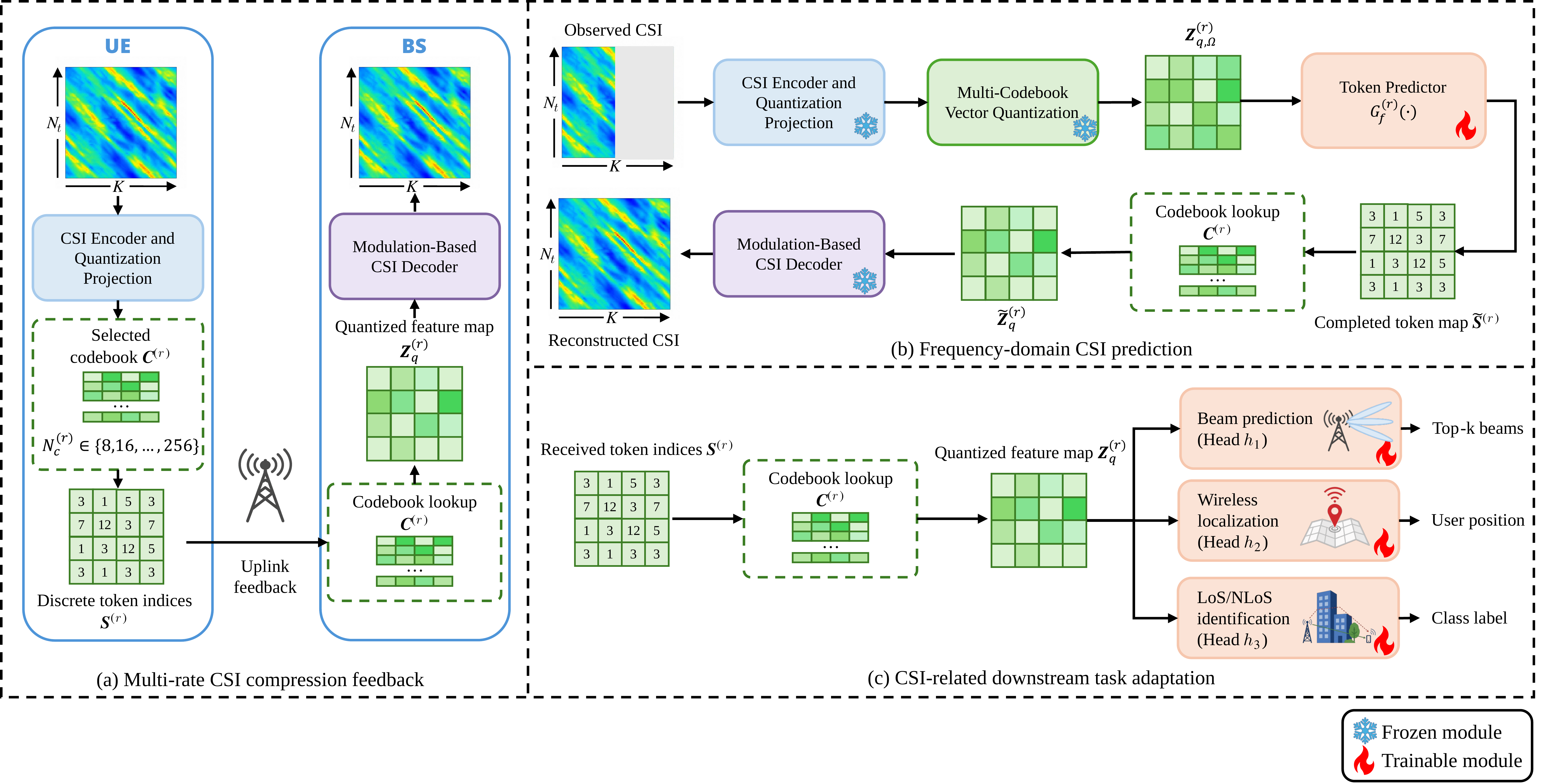}
\caption{Task adaptation of WiFo-D for multi-rate CSI compression
feedback, frequency-domain CSI prediction, and CSI-related
downstream tasks.}
\label{fig:task_adaptation}
\end{figure*}

This section presents how WiFo-D applies the learned discrete
CSI representations to multi-rate CSI feedback,
frequency-domain CSI prediction, and CSI-related downstream
tasks, as illustrated in Fig.~\ref{fig:task_adaptation}. For
CSI feedback, the pre-trained mappings $F_{\rm e}(\cdot)$,
$Q^{(r)}(\cdot)$, and $F_{\rm d}(\cdot)$ are directly used
under different codebook choices. For frequency-domain
prediction, these mappings remain frozen and only a lightweight
token predictor is trained. For downstream tasks, lightweight
task heads are optimized using the quantized feature maps.

\subsection{Multi-Rate CSI Compression Feedback}

As illustrated in Fig.~\ref{fig:task_adaptation}(a), WiFo-D
uses discrete token indices as the finite-rate feedback message.
Unlike continuous-feature feedback methods that require an
additional quantization stage~\cite{ref_TransNet,ref_WiFo_CF},
the UE maps the downlink CSI to discrete token indices through
$F_{\rm e}(\cdot)$ and $Q^{(r)}(\cdot)$ under the selected
codebook:
\begin{equation}
\left(
\mathbf{S}^{(r)},
\mathbf{Z}_{\rm q}^{(r)}
\right)
=
Q^{(r)}
\left(
F_{\rm e}(\mathbf{H});
\mathcal{C}^{(r)}
\right).
\end{equation}
Only the discrete token map $\mathbf{S}^{(r)}$ is transmitted
through the feedback link. After receiving the token indices,
the BS recovers the quantized feature map as
\begin{equation}
\mathbf{Z}_{\rm q}^{(r)}
=
\operatorname{Lookup}
\left(
\mathbf{S}^{(r)};
\mathcal{C}^{(r)}
\right),
\end{equation}
and reconstructs the CSI through
\begin{equation}
\hat{\mathbf{H}}^{(r)}
=
F_{\rm d}
\left(
\mathbf{Z}_{\rm q}^{(r)}
\right).
\end{equation}

For a token map containing $L=H_zW_z$ positions, the feedback
bits and normalized overhead are
\begin{equation}
\begin{gathered}
N_{\rm bit}^{(r)}
=
L
\left\lceil
\log_2 N_{\rm c}^{(r)}
\right\rceil,\\[3pt]
\rho_{\rm fb}^{(r)}
=
\frac{N_{\rm bit}^{(r)}}{2N_tK}.
\end{gathered}
\end{equation}
Each token index selects a learned
$d_{\rm c}$-dimensional channel pattern. Smaller codebooks
provide compact coarse-grained feedback, whereas larger
codebooks provide finer-grained representations at higher
overhead. Since all codebooks share the same encoder-decoder
backbone, the feedback rate is changed by selecting a codebook
rather than training a separate model.

\subsection{Discrete-Token-Based Frequency-Domain CSI Prediction}

As illustrated in Fig.~\ref{fig:task_adaptation}(b), WiFo-D formulates frequency-domain CSI prediction as
conditional generation of a complete token map over the
learned codebook, drawing
on the discrete representation and token-prediction paradigms
in~\cite{ref_VQGAN,ref_MaskGIT,ref_BEiT,ref_Emu3}.
Following the formulation in Section II-B, $\Omega$ and
$\bar{\Omega}$ denote the observed and unobserved frequency
index sets, respectively. The partially observed CSI
$\mathbf{H}_{\Omega}$ retains the complete antenna-domain
channel vectors over $\Omega$ and masks the entire frequency
columns over $\bar{\Omega}$. Given
$\mathbf{H}_{\Omega}$, the frozen mappings $F_{\rm e}(\cdot)$ and
$Q^{(r)}(\cdot)$ first produce
\begin{equation}
\left(
\mathbf{S}_{\Omega}^{(r)},
\mathbf{Z}_{{\rm q},\Omega}^{(r)}
\right)
=
Q^{(r)}
\left(
F_{\rm e}
\left(
\mathbf{H}_{\Omega}
\right);
\mathcal{C}^{(r)}
\right).
\end{equation}
The lightweight predictor then takes the observed quantized
feature map as input:
\begin{equation}
\mathbf{P}^{(r)}
=
G_{\rm f}^{(r)}
\left(
\mathbf{Z}_{{\rm q},\Omega}^{(r)}
\right),
\end{equation}
where $\mathbf{P}^{(r)}$ contains the predicted codeword
probabilities over all positions of the complete token map.
The observed token map $\mathbf{S}_{\Omega}^{(r)}$ is not used
as the predictor input.

During inference, the complete token map is obtained by
\begin{equation}
\widetilde{\mathbf{S}}^{(r)}(i)
=
\arg\max_{1\leq j\leq N_{\rm c}^{(r)}}
\left[
\mathbf{p}_i^{(r)}
\right]_j,
\qquad
i\in\mathcal{I},
\end{equation}
where $\mathcal{I}$ denotes all positions in the complete token
map. The predicted token map is converted into the corresponding
quantized feature map, from which the CSI is reconstructed as
\begin{equation}
\begin{gathered}
\widetilde{\mathbf{Z}}_{\rm q}^{(r)}
=
\operatorname{Lookup}
\left(
\widetilde{\mathbf{S}}^{(r)};
\mathcal{C}^{(r)}
\right),\\[3pt]
\widetilde{\mathbf{H}}^{(r)}
=
F_{\rm d}
\left(
\widetilde{\mathbf{Z}}_{\rm q}^{(r)}
\right).
\end{gathered}
\end{equation}

For training, the complete CSI is processed by the frozen
mappings $F_{\rm e}(\cdot)$ and $Q^{(r)}(\cdot)$ to obtain
the target token map $\mathbf{S}^{(r)}$. The token prediction
loss is
\begin{equation}
\ell_{\rm tok}^{(r)}
=
-
\frac{1}{|\mathcal{I}|}
\sum_{i\in\mathcal{I}}
\log
\left(
\left[
\mathbf{p}_i^{(r)}
\right]_{s_i^{(r)}}
\right),
\end{equation}
where $s_i^{(r)}$ is the ground-truth codeword index at position
$i$.

To provide differentiable CSI-domain supervision, the predicted
probabilities are converted into a soft quantized feature at each
token position:
\begin{equation}
\overline{\mathbf{z}}_{{\rm q},i}^{(r)}
=
\sum_{j=1}^{N_{\rm c}^{(r)}}
\left[
\mathbf{p}_i^{(r)}
\right]_j
\mathbf{e}_j^{(r)}.
\end{equation}
Collecting the soft features gives
$\overline{\mathbf{Z}}_{\rm q}^{(r)}$, which is decoded as
\begin{equation}
\widetilde{\mathbf{X}}^{(r)}
=
D_{\psi}
\left(
\mathbf{F}_0;
\overline{\mathbf{Z}}_{\rm q}^{(r)}
\right).
\end{equation}
With $\mathbf{X}=\mathcal{T}(\mathbf{H})$, the reconstruction
loss over the missing frequency region is
\begin{equation}
\ell_{\rm gen}^{(r)}
=
\frac{
\left\|
\mathbf{A}_{\bar{\Omega}}
\odot
\left(
\widetilde{\mathbf{X}}^{(r)}
-
\mathbf{X}
\right)
\right\|_F^2
}{
\left\|
\mathbf{A}_{\bar{\Omega}}
\right\|_1
}.
\end{equation}
The overall frequency-prediction objective is
\begin{equation}
\ell_{\rm f}^{(r)}
=
\lambda_{\rm tok}
\ell_{\rm tok}^{(r)}
+
\lambda_{\rm gen}
\ell_{\rm gen}^{(r)},
\end{equation}
where $\mathbf{A}_{\bar{\Omega}}$ is the missing-frequency
mask, and $\lambda_{\rm tok}$ and $\lambda_{\rm gen}$ are loss
weights. During task adaptation, only
$G_{\rm f}^{(r)}(\cdot)$ is optimized, while
$F_{\rm e}(\cdot)$, $Q^{(r)}(\cdot)$, and
$F_{\rm d}(\cdot)$ remain frozen.

\subsection{CSI-Related Downstream Task Adaptation}

As illustrated in Fig.~\ref{fig:task_adaptation}(c), the
quantized feature map can be directly used for downstream
inference without first reconstructing continuous CSI. When the
token indices are available at the BS, the corresponding feature
map is recovered by
\begin{equation}
\mathbf{Z}_{\rm q}^{(r)}
=
\operatorname{Lookup}
\left(
\mathbf{S}^{(r)};
\mathcal{C}^{(r)}
\right).
\end{equation}
During task adaptation, $F_{\rm e}(\cdot)$ and
$Q^{(r)}(\cdot)$ remain frozen, and only lightweight task
heads are optimized.

We consider sub-6 GHz-to-mmWave beam prediction, wireless
localization, and LoS/NLoS identification
~\cite{ref_DL_BP,ref_WiFo2}. The beam-prediction head $h_1$
maps the sub-6 GHz quantized feature map to a probability
distribution over candidate mmWave beams, the localization
head $h_2$ estimates the UE position, and the LoS/NLoS
identification head $h_3$ outputs the corresponding class
probabilities. The beam-prediction and LoS/NLoS identification
heads are trained with cross-entropy losses, whereas the
localization head uses mean-square error. The same quantized representation can also be
decoded by $F_{\rm d}(\cdot)$ for a reconstruct-then-infer
pipeline, which is evaluated in Section V.

\section{Experiments}

This section evaluates WiFo-D on multi-rate CSI feedback,
frequency-domain CSI prediction, and downstream task adaptation.
Ablation, scaling, and complexity studies further examine the
key designs and deployment efficiency.

\subsection{Experimental Setup}

\subsubsection{Dataset Construction and Splits}
We construct 27 heterogeneous CSI datasets using
QuaDRiGa-based stochastic simulations~\cite{ref_QuaDRiGa}
and DeepMIMO-based ray-tracing channels~\cite{ref_DeepMIMO}.
The QuaDRiGa datasets cover indoor, urban microcell (UMi),
urban macrocell (UMa), and rural macrocell (RMa)
environments under both LoS and NLoS conditions, while the
DeepMIMO datasets include the I1, I3, and O1 scenarios.
Across these datasets, the carrier frequency ranges from
2.4 to 6.7 GHz, the number of subcarriers is selected from
$\{16,32,64\}$, and the subcarrier spacing is selected from
$\{15,30,60\}$ kHz. Different planar-array configurations are
also included to increase configuration diversity.

Each dataset contains 15,000 independently generated CSI
samples, divided into 10,000 training, 2,000 validation, and
3,000 test samples. For D1--D22, the training samples are
used for pre-training, while the corresponding validation and
test samples are used for model selection and ID evaluation,
respectively. D23--D27 are excluded from WiFo-D pre-training
and reserved for OOD evaluation. Specifically, D23--D25
introduce carrier frequencies of 2.8, 5.3, and 6.2 GHz,
respectively, which are absent from the pre-training datasets.
D24 additionally adopts an unseen frequency-sampling
configuration with $K=64$ and $\Delta f=30$ kHz. D26 and
D27 use previously observed DeepMIMO scenarios and parameter
values but form unseen scenario--array combinations, namely
I1\_2p5 with a $32\times1$ array and O1\_3p4 with a
$16\times2$ array. The OOD datasets therefore cover both
unseen parameter values and unseen combinations of propagation
and system configurations. Each CSI
sample is independently normalized after real-imaginary
stacking. A compact overview of the dataset splits and
configuration ranges is provided in
Table~\ref{tab:dataset_config}.

\begin{table*}[t]
\centering
\caption{Overview of pre-training and evaluation datasets.}
\label{tab:dataset_config}
\small
\renewcommand{\arraystretch}{1.2}

\begin{tabular*}{\textwidth}{
@{\extracolsep{\fill}}
c c c c c c c c
@{}
}
\noalign{\hrule height 1.0pt}
Set type &
Source &
Dataset IDs &
$f_c$ (GHz) &
$N_{\rm h}$ &
$N_{\rm v}$ &
$K$ &
Samples (k) \\
\noalign{\hrule height 0.4pt}

\multirow{2}{*}{\shortstack{Pre-training/ID}}
& QuaDRiGa
& D1--D16
& 2.4--6.7
& 1--8
& 8--32
& 16/32/64
& 240 \\

& DeepMIMO
& D17--D22
& 2.4--3.5
& 8--32
& 1--8
& 32/64
& 90 \\

\noalign{\hrule height 0.4pt}

\multirow{2}{*}{OOD}
& QuaDRiGa
& D23--D25
& 2.8--6.2
& 2--8
& 8--16
& 32/64
& 45 \\

& DeepMIMO
& D26--D27
& 2.5--3.4
& 16--32
& 1--2
& 32/64
& 30 \\

\noalign{\hrule height 1.0pt}
\end{tabular*}
\end{table*}

\subsubsection{Baselines}
For CSI feedback, we select four representative baselines
covering complementary paradigms. TVAL3~\cite{ref_TVAL3} is
an optimization-based method that reconstructs CSI using a
total-variation prior and does not contain learned parameters.
TransNet~\cite{ref_TransNet} is a strong task-specific
deep-learning feedback model based on full attention. SG-VQ
~\cite{ref_VQ_CSI_Feedback} is a task-specific
vector-quantized feedback model that represents latent features
using finite-bit codeword indices. WiFo-CF
~\cite{ref_WiFo_CF} is a wireless foundation-model baseline
that jointly supports heterogeneous CSI feedback based on
continuous representations. Together, these methods cover
model-driven optimization, task-specific deep learning,
task-specific vector quantization, and foundation-model-based
feedback.

For frequency-domain prediction, Transformer-CP
~\cite{ref_Transformer_CP} serves as a direct CSI prediction
baseline, while the continuous counterpart WiFo-D-Cont
performs continuous latent prediction under the same backbone
to isolate the effect of discretization. For beam prediction,
localization, and LoS/NLoS identification, the task-specific
baselines are BP-DNN~\cite{ref_DL_BP}, WiT~\cite{ref_WiT},
and ST-CNN~\cite{ref_STCNN}, respectively. All baselines
follow their original training settings with necessary
input-format adjustments.

\subsubsection{Evaluation Metrics}
For CSI reconstruction-oriented tasks, including CSI compression
feedback and frequency-domain CSI prediction, we use the
normalized mean square error (NMSE) as the main performance
metric. Given the reconstructed or predicted CSI
$\hat{\mathbf{H}}$ and the ground-truth CSI $\mathbf{H}$, NMSE
is defined as
\begin{equation}
{\rm NMSE}
=
\mathbb{E}
\left\{
\frac{
\left\|
\hat{\mathbf{H}}
-
\mathbf{H}
\right\|_F^2
}{
\left\|
\mathbf{H}
\right\|_F^2
}
\right\}.
\end{equation}
Unless otherwise specified, we report NMSE in dB, i.e.,
$10\log_{10}({\rm NMSE})$. For frequency-domain CSI
prediction, NMSE is computed only over the unobserved frequency
region $\bar{\Omega}$. For codebook analysis, we report the
codebook usage ratio and normalized perplexity. For downstream
tasks, Top-$k$ accuracy is used for sub-6 GHz-to-mmWave beam
prediction, root mean square error (RMSE) is used for wireless localization, and
classification accuracy is used for LoS/NLoS identification.
Model complexity is evaluated in terms of the number of
parameters, floating-point operations (FLOPs), inference latency,
and model storage.

\subsubsection{Implementation Details}
Unless otherwise specified, experiments use the configuration in
Table~\ref{tab:wifod_config}. Six codebooks are jointly trained
with uniform sampling to support token widths from 3 to 8 bits.

\begin{table}[!tbp]
\centering
\caption{Default network and training configuration of WiFo-D.}
\label{tab:wifod_config}
\small
\renewcommand{\arraystretch}{1.2}
\setlength{\tabcolsep}{8pt}
\begin{tabular}{c c}
\noalign{\hrule height 1.0pt}
\textbf{Parameter} & \textbf{Value} \\
\noalign{\hrule height 0.4pt}
Base channels & 128 \\
Channel multipliers & $[1,2,2]$ \\
Latent/codeword dimension & 4 \\
Latent downsampling factor & $4\times4$ \\
Fourier bands & 6 \\
Codebook sizes & $\{256,128,64,32,16,8\}$ \\
Codebook sampling & Uniform \\
Optimizer & AdamW \\
Batch size & 64 \\
Learning rate / weight decay & $2\times10^{-4}$ / $10^{-2}$ \\
Rec./codebook/commitment weights & $1.0/1.0/0.25$ \\
\noalign{\hrule height 1.0pt}
\end{tabular}
\end{table}

\subsection{Multi-Rate CSI Feedback}

We first evaluate WiFo-D under 3-bit ID and OOD feedback and
then analyze its multi-rate behavior and codebook utilization.

\subsubsection{3-bit In-Distribution CSI Feedback}
We use a latent compression ratio of 1/32 and impose the
same total feedback-bit budget on all learned feedback
methods. With the $4\times4$ latent downsampling factor,
WiFo-D produces
$L=H_zW_z=N_tK/16$ token positions. Under the 3-bit
setting, $N_{\rm c}=8$, and each token index requires
$\log_2N_{\rm c}=3$ bits, resulting in
$N_{\rm bit}^{\rm WiFo-D}=3L=3N_tK/16$.

In our implementations, TransNet and WiFo-CF each produce a
length-$L$ continuous feedback vector, quantized element-wise
using 3-bit uniform scalar quantization and the default 3-bit
$\mu$-law scalar quantization, respectively, both yielding
$3L$ feedback bits. SG-VQ uses $L$ latent positions, with
2 bits for the shape index and 1 bit for the gain per position,
also yielding $3L$ bits. Thus, all learned methods use the same
total feedback-bit budget but different feedback representations.
The latent compression ratio is $L/(2N_tK)=1/32$, and the
normalized feedback-bit overhead defined in Section IV-A is
$\rho_{\rm fb}=3L/(2N_tK)=3/32$.

TransNet and SG-VQ are trained separately for each domain,
whereas WiFo-CF and WiFo-D use one jointly trained model
across all ID datasets.

\begin{table}[!tbp]
\centering
\caption{3-bit ID CSI feedback NMSE (dB) at compression ratio $1/32$.}
\label{tab:id_feedback_3bit}
\small
\renewcommand{\arraystretch}{1.2}
\setlength{\tabcolsep}{5.3pt}
\begin{tabular}{c|c c c c c}
\noalign{\hrule height 1.0pt}
\textbf{Dataset} & \textbf{WiFo-D} & \textbf{TVAL3} & \textbf{TransNet} & \textbf{SG-VQ} & \textbf{WiFo-CF} \\
\noalign{\hrule height 0.4pt}
D1  & \textbf{-20.433} & -1.033 & -10.240 & -9.143 & \underline{-12.599} \\
D2  & \textbf{-14.823} & -0.265 & \underline{-9.601}  & -7.180 & -7.512 \\
D3  & \textbf{-15.937} & -0.840 & \underline{-11.183} & -10.848 & -10.677 \\
D4  & \textbf{-13.748} & -0.936 & -7.499 & \underline{-8.020} & -7.696 \\
D5  & \textbf{-15.264} & -0.718 & \underline{-9.828} & -7.872 & -9.803 \\
D6  & \textbf{-11.996} & -0.760 & -6.499 & \underline{-7.178} & -7.076 \\
D7  & \textbf{-19.443} & -2.341 & -11.376 & -10.936 & \underline{-12.292} \\
D8  & \textbf{-20.848} & -1.468 & -12.539 & -8.314 & \underline{-13.464} \\
D9  & \textbf{-20.988} & -0.803 & \underline{-15.160} & -6.549 & -13.545 \\
D10 & \textbf{-9.413}  & -0.223 & \underline{-6.936} & -6.370 & -2.769 \\
D11 & \textbf{-22.134} & -1.755 & \underline{-13.933} & -8.935 & -11.748 \\
D12 & \textbf{-14.129} & -1.351 & \underline{-8.803} & -8.309 & -8.494 \\
D13 & \textbf{-17.406} & -0.941 & \underline{-10.006} & -7.166 & -8.167 \\
D14 & \textbf{-9.632}  & -0.750 & -5.398 & \underline{-6.029} & -5.776 \\
D15 & \textbf{-17.730} & -0.317 & -10.375 & -8.734 & \underline{-11.001} \\
D16 & \textbf{-20.469} & -1.076 & -11.736 & -7.731 & \underline{-11.911} \\
D17 & \textbf{-19.687} & 0.122  & \underline{-15.947} & -11.183 & -11.804 \\
D18 & \textbf{-18.641} & 0.110  & \underline{-16.720} & -10.862 & -11.815 \\
D19 & \textbf{-21.380} & -0.817 & \underline{-18.538} & -12.372 & -11.372 \\
D20 & \textbf{-22.647} & 0.115  & \underline{-15.110} & -12.457 & -14.997 \\
D21 & \textbf{-21.635} & 0.115  & \underline{-17.218} & -11.246 & -15.036 \\
D22 & \textbf{-23.863} & 0.125  & \underline{-16.481} & -11.932 & -15.372 \\
\noalign{\hrule height 0.4pt}
\textbf{Avg.} & \textbf{-15.713} & -0.672 & \underline{-10.414} & -8.625 & -9.391 \\
\noalign{\hrule height 1.0pt}
\end{tabular}
\end{table}

As shown in Table~\ref{tab:id_feedback_3bit}, WiFo-D
achieves the best NMSE on all ID datasets, with an average
of -15.713 dB. It outperforms TransNet, the strongest
task-specific baseline, by 5.299 dB on average, although
TransNet is trained separately for each domain. Despite using
only 3 bits per token and a compact codebook with
$N_{\rm c}=8$, WiFo-D remains effective because multiple
spatial-frequency token positions jointly form a combinatorial
CSI representation. Each codeword captures a local CSI
pattern, while the modulation-based decoder supplies a
structural reconstruction prior learned from heterogeneous
CSI data. WiFo-D therefore reconstructs CSI from compact
channel-pattern indices without transmitting high-precision
continuous features.

TVAL3 is limited by its handcrafted total-variation prior.
TransNet and SG-VQ require separate domain-specific models,
and SG-VQ cannot benefit from heterogeneous pre-training. WiFo-CF uses joint foundation-model training
and ranks second on five ID datasets, but its average NMSE
is -9.391 dB. WiFo-D further improves this result by
6.322 dB, demonstrating the advantage of its native discrete
feedback interface under stringent feedback budgets.

\subsubsection{3-bit Out-of-Distribution Generalization}
We further evaluate OOD generalization under the same 3-bit
feedback setting and total feedback-bit budget, with a latent
compression ratio of 1/32.
The OOD datasets D23--D27 are excluded from the training data
of both WiFo-D and WiFo-CF. The two foundation models use neither the training nor
validation samples from these domains and are directly
evaluated on the common 3,000-sample OOD test sets, thereby
constituting zero-shot methods. TransNet
and SG-VQ are separately trained on the 10,000 training samples
of each target OOD domain, selected using the corresponding
2,000 validation samples, and evaluated on the same test sets.
They therefore serve as full-shot task-specific baselines.
TVAL3 contains no learned parameters and is directly applied
to each target domain. This setting evaluates whether the learned representations can
transfer to unseen carrier frequencies and frequency-sampling
configurations, as well as unseen combinations of propagation
scenarios and antenna configurations. The superscript ``ZS'' in
Table~\ref{tab:ood_feedback_3bit} denotes zero-shot evaluation.

\begin{table}[!tbp]
\centering
\caption{3-bit OOD CSI feedback NMSE (dB) at compression ratio $1/32$.}
\label{tab:ood_feedback_3bit}
\small
\renewcommand{\arraystretch}{1.2}
\setlength{\tabcolsep}{3.8pt}
\begin{tabular}{c|c c c c c}
\noalign{\hrule height 1.0pt}
\textbf{Dataset} & \textbf{WiFo-D$^{\rm ZS}$} &
\textbf{TVAL3} & \textbf{TransNet} & \textbf{SG-VQ} &
\textbf{WiFo-CF$^{\rm ZS}$} \\
\noalign{\hrule height 0.4pt}
D23 & \textbf{-14.786} & -0.423 & \underline{-9.792}  & -7.289  & -7.494 \\
D24 & \textbf{-21.039} & -1.899 & \underline{-12.520} & -9.343  & -10.908 \\
D25 & \textbf{-10.896} & -0.947 & -6.599 & \underline{-6.854} & -5.925 \\
D26 & \textbf{-19.087} & 0.123  & \underline{-15.517} & -13.000 & -11.282 \\
D27 & \textbf{-20.618} & 0.119  & \underline{-17.935} & -11.388 & -14.494 \\
\noalign{\hrule height 0.4pt}
\textbf{Avg.} & \textbf{-15.422} & -0.542 & \underline{-10.717} & -8.982 & -9.033 \\
\noalign{\hrule height 1.0pt}
\end{tabular}
\end{table}

Table~\ref{tab:ood_feedback_3bit} shows that WiFo-D ranks
first on all OOD datasets and achieves an average NMSE of
-15.422 dB. Despite using no target-domain training samples,
WiFo-D outperforms the full-shot TransNet baseline by 4.705 dB
on average. It also consistently outperforms the zero-shot
WiFo-CF baseline, whose average NMSE is -9.033 dB. These
results indicate that the learned discrete codebooks transfer
effectively across held-out propagation and system
configurations.

\subsubsection{Multi-Rate Feedback Behavior}
\begin{figure}[!tbp]
\centering
\includegraphics[width=0.90\linewidth]{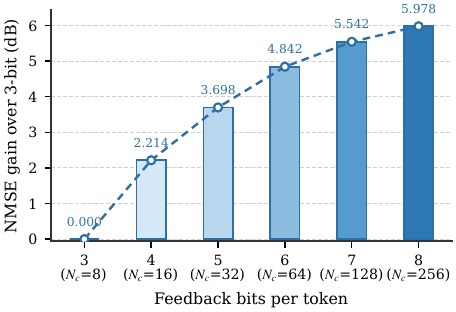}
\caption{Multi-rate CSI feedback performance of WiFo-D on the ID datasets.}
\label{fig:b3_multirate}
\end{figure}

We then evaluate the multi-rate behavior of WiFo-D on the ID datasets by varying the codebook size from $N_{\rm c}=8$ to $N_{\rm c}=256$, corresponding to 3-bit to 8-bit token indices. As shown in Fig.~\ref{fig:b3_multirate}, increasing the feedback bit width consistently improves the reconstruction performance. Compared with the 3-bit setting, the 8-bit setting with $N_{\rm c}=256$ achieves a 5.978 dB NMSE gain, improving the average NMSE from -15.713 dB to -21.691 dB.

The performance gain gradually saturates as the codebook size increases. This trend is consistent with the expected coarse-to-fine behavior of the learned discrete representation. A small codebook provides compact but coarse CSI tokens for stringent feedback budgets, while a larger codebook offers finer-grained channel patterns for more accurate reconstruction. Importantly, all rates are supported by the same WiFo-D model through jointly trained multi-rate codebooks, rather than requiring separate feedback models for different bit budgets. This result shows that WiFo-D provides a flexible discrete feedback interface that can adapt to different uplink feedback constraints.

\subsubsection{Codebook Usage Analysis}
We examine codebook utilization using the usage ratio and
normalized perplexity. The usage ratio is the percentage of
codewords activated at least once on the ID test samples and
therefore reflects the prevalence of dead codewords. Given
the empirical probability $p_i$ of the $i$-th codeword, the
normalized perplexity is
$\exp(-\sum_i p_i\log p_i)/N_{\rm c}$, expressed as a
percentage, and measures the effective diversity of the token
distribution.

\begin{figure}[!tbp]
\centering
\includegraphics[width=0.96\linewidth]{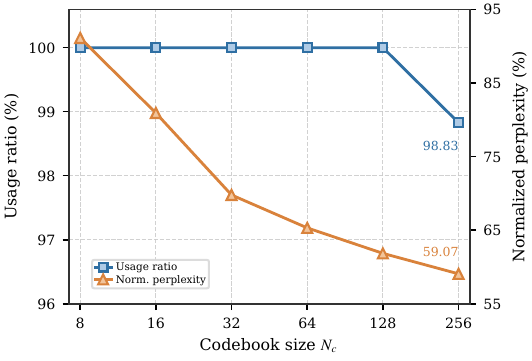}
\caption{Codebook usage analysis of WiFo-D under different codebook sizes.}
\label{fig:codebook_usage}
\end{figure}

As shown in Fig.~\ref{fig:codebook_usage}, all codewords are
activated for $N_{\rm c}\leq128$, while the usage ratio remains
98.83\% for $N_{\rm c}=256$, indicating no substantial
codebook collapse. The normalized perplexity decreases with
the codebook size and reaches 59.07\% at $N_{\rm c}=256$,
revealing a more long-tailed distribution in which frequent
codewords represent common CSI patterns and rarer codewords
capture finer structures. The high usage ratio and non-trivial
perplexity confirm that WiFo-D learns a diverse discrete CSI
vocabulary supporting its low-bit and multi-rate feedback
performance.

\subsection{Frequency-Domain CSI Prediction}

We compare discrete-token prediction with direct CSI prediction
and continuous latent prediction under limited-data,
cross-domain, and noisy-observation settings.

\subsubsection{Prediction Baselines and Settings}
Transformer-CP directly predicts the missing CSI from the
observed frequency samples. To isolate the effect of
discretization, WiFo-D-Cont removes the VQ module and
codebooks while retaining the same encoder-decoder backbone,
which is pre-trained using CSI reconstruction. During
prediction, its backbone is frozen and only a continuous
predictor mapping the observed latent representation to the
complete one is trained. WiFo-D instead freezes the
pre-trained encoder, codebook, and decoder and trains only
the token predictor.

Training the same token-prediction pipeline from scratch
produces substantially worse NMSE, confirming that the
pre-trained discrete CSI vocabulary is essential. This variant
is therefore omitted from the main comparison. During frequency-domain prediction, Transformer-CP,
WiFo-D-Cont, and WiFo-D contain 1.887M, 0.746M, and
0.749M trainable parameters, respectively. The frozen
pre-trained modules of the latter two methods are excluded
from these counts.

To avoid overlap with the pre-training datasets in
Table~\ref{tab:dataset_config}, we construct two additional
domains, FP1 and FP2. Unless otherwise specified, experiments
use FP1, an RMa LoS domain with $f_c=7.0$ GHz, $K=64$,
$\Delta f=15$ kHz, $B=0.96$ MHz, and
$N_{\rm h}\times N_{\rm v}=2\times32$ ($N_t=64$).
The observed-frequency ratio specifies the fraction of
available frequency samples, and performance is measured by
the NMSE over the remaining region.

\subsubsection{Sample Efficiency under Limited Observations}
\begin{figure}[!tbp]
\centering
\includegraphics[width=0.94\linewidth]{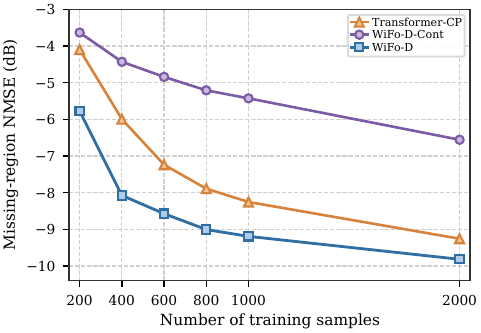}
\caption{Sample efficiency for frequency-domain CSI prediction.}
\label{fig:freq_sample_efficiency}
\end{figure}

We first set the observed-frequency ratio to 0.25.
As shown in Fig.~\ref{fig:freq_sample_efficiency}, WiFo-D
outperforms Transformer-CP and WiFo-D-Cont at every training
scale. With 200 samples, it achieves -5.776 dB, improving
over the two baselines by 1.676 dB and 2.141 dB,
respectively. With 1000 samples, it remains best at
-9.191 dB. Direct CSI prediction must learn a continuous
mapping to the missing coefficients, while continuous latent
prediction still operates in an unconstrained feature space.
By predicting channel-pattern indices from a pre-trained
vocabulary, WiFo-D obtains a more structured target and
better sample efficiency.

\subsubsection{Representative Cross-Domain Transfer}
We next consider transfer from FP1 to FP2, a UMa LoS domain
with $f_c=6.1$ GHz, $K=64$, $\Delta f=30$ kHz,
$B=1.92$ MHz, and
$N_{\rm h}\times N_{\rm v}=2\times32$ ($N_t=64$).
WiFo-D and WiFo-D-Cont are trained on FP1 and evaluated
zero-shot on FP2, whereas Transformer-CP is fully trained on
FP2. As shown in Table~\ref{tab:freq_cross_domain}, zero-shot
WiFo-D achieves -4.714 dB, close to the -4.747 dB of
full-shot Transformer-CP, and outperforms zero-shot
WiFo-D-Cont by 1.464 dB. Although this representative case
does not establish universal OOD superiority, it demonstrates
that the discrete vocabulary can transfer to a held-out
prediction domain and provides a more transferable intermediate
representation than continuous latent features.

\begin{table}[!tbp]
\centering
\caption{Cross-domain frequency prediction from FP1 to FP2.}
\label{tab:freq_cross_domain}
\small
\renewcommand{\arraystretch}{1.2}
\setlength{\tabcolsep}{6pt}
\begin{tabular}{l c c}
\noalign{\hrule height 1.0pt}
\textbf{Method} & \textbf{Training protocol} & \textbf{NMSE (dB)} \\
\noalign{\hrule height 0.4pt}
Transformer-CP & FP2 full-shot & \textbf{-4.747} \\
WiFo-D & FP1-to-FP2 zero-shot & \underline{-4.714} \\
WiFo-D-Cont & FP1-to-FP2 zero-shot & -3.250 \\
\noalign{\hrule height 1.0pt}
\end{tabular}
\end{table}

\subsubsection{Robustness to Noisy Observations}
We corrupt the observed FP1 samples with additive noise while
retaining clean CSI as the missing-region target. WiFo-D and
WiFo-D-Cont are trained with SNR uniformly sampled from
-5 to 20 dB, enabling a controlled comparison of discrete
and continuous representations under the same framework.
As shown in Table~\ref{tab:freq_snr_robustness}, WiFo-D
outperforms WiFo-D-Cont at every test SNR. Its gain increases
from 0.750 dB at -5 dB to 3.281 dB at 20 dB, indicating
that token prediction suppresses noise-induced continuous
perturbations and yields a more robust intermediate
representation.

\subsection{CSI-Related Downstream Task Adaptation}

We further evaluate whether the discrete representations learned by WiFo-D can be transferred to CSI-related downstream tasks. Different from CSI reconstruction-oriented evaluations, this subsection focuses on whether the quantized features obtained from the frozen WiFo-D encoder and codebook preserve task-relevant channel information. Three representative tasks are considered, including sub-6 GHz-to-mmWave beam prediction, wireless localization, and LoS/NLoS identification. The corresponding task-specific baselines are BP-DNN, WiT, and ST-CNN, respectively. The downstream datasets used in this subsection are held out from WiFo-D pre-training.

\subsubsection{Overall Downstream Task Performance}
Table~\ref{tab:downstream_adaptation} summarizes the downstream task adaptation results. For WiFo-D, the pre-trained encoder and codebook are frozen, and only a lightweight task head is trained on the quantized feature map. As shown in Table~\ref{tab:downstream_adaptation}, WiFo-D consistently outperforms the task-specific baselines across all three downstream tasks.

For beam prediction, WiFo-D improves the Top-1 accuracy from 82.17\% to 85.80\%, and achieves more significant gains in Top-3 and Top-5 accuracy. This indicates that the quantized features preserve beam-relevant spatial channel structures. For wireless localization, WiFo-D reduces the RMSE from 1.126 m to 0.794 m, showing that the learned discrete representation also contains useful position-related propagation information. For LoS/NLoS identification, WiFo-D improves the classification accuracy from 78.98\% to 88.67\%, demonstrating that the quantized features can capture propagation-environment-related characteristics. These results show that WiFo-D does not merely serve as a CSI reconstruction model, but provides a compact and transferable representation interface for heterogeneous downstream wireless tasks.

\begin{table}[!tbp]
\centering
\caption{Robustness to noisy observations on FP1.}
\label{tab:freq_snr_robustness}
\small
\renewcommand{\arraystretch}{1.2}
\setlength{\tabcolsep}{5.3pt}
\begin{tabular}{l c c c c c}
\noalign{\hrule height 1.0pt}
\textbf{Method} & \textbf{-5 dB} & \textbf{0 dB} & \textbf{5 dB} & \textbf{10 dB} & \textbf{20 dB} \\
\noalign{\hrule height 0.4pt}
WiFo-D & \textbf{-3.495} & \textbf{-5.783} & \textbf{-7.214} & \textbf{-7.931} & \textbf{-8.344} \\
WiFo-D-Cont & -2.745 & -3.799 & -4.507 & -4.882 & -5.063 \\
\noalign{\hrule height 0.4pt}
Gain over Cont. & 0.750 & 1.984 & 2.707 & 3.049 & 3.281 \\
\noalign{\hrule height 1.0pt}
\end{tabular}
\end{table}

\begin{table*}[!t]
\centering
\caption{Downstream task adaptation results on held-out datasets
unseen during WiFo-D pre-training.}
\label{tab:downstream_adaptation}
\small
\renewcommand{\arraystretch}{1.3}

\begin{tabular*}{0.96\textwidth}{
@{\extracolsep{\fill}}
c c c c c
@{}
}
\noalign{\hrule height 1.0pt}
\textbf{Task} &
\textbf{Metric} &
\textbf{Baseline} &
\textbf{Baseline result} &
\textbf{WiFo-D} \\
\noalign{\hrule height 0.4pt}

Beam prediction &
Top-1 / Top-3 / Top-5 Acc. (\%) &
BP-DNN &
82.17 / 90.37 / 92.37 &
\textbf{85.80 / 97.80 / 98.87} \\

Wireless localization &
RMSE (m) &
WiT &
1.126 &
\textbf{0.794} \\

LoS/NLoS identification &
Acc. (\%) &
ST-CNN &
78.98 &
\textbf{88.67} \\

\noalign{\hrule height 1.0pt}
\end{tabular*}
\end{table*}

\subsubsection{Inference Pipeline Analysis}
To further understand the relationship between CSI reconstruction and downstream inference, we conduct a representative pipeline analysis on wireless localization with WiT. Four settings are compared, including raw CSI fed into WiT, decoded CSI fed into WiT, a raw-CSI-trained WiT directly evaluated on decoded CSI, and WiFo-D quantized features fed into a lightweight task head.

\begin{figure}[!tbp]
\centering
\includegraphics[width=0.90\linewidth]{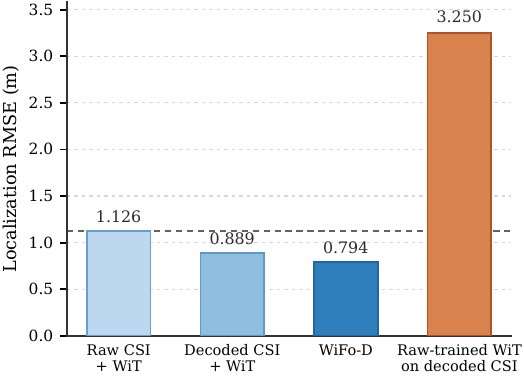}
\caption{Localization performance with different inference pipelines.}
\label{fig:downstream_pipeline}
\end{figure}

As shown in Fig.~\ref{fig:downstream_pipeline}, decoded CSI achieves a lower localization RMSE than raw CSI when WiT is trained and evaluated on the corresponding input domain. This suggests that WiFo-D reconstruction preserves task-relevant channel structures and can act as a structured CSI projection for downstream training. However, when a WiT model trained on raw CSI is directly evaluated on decoded CSI, the RMSE increases sharply to 3.250 m. This indicates that reconstructed CSI and raw CSI may follow different feature distributions, so a task model trained on raw CSI cannot always be directly reused after CSI reconstruction.

In contrast, the WiFo-D quantized feature achieves the best localization RMSE of 0.794 m with a lightweight task head. This result suggests that, although reconstructed CSI can preserve useful structures, the quantized feature provides a more direct and efficient downstream adaptation interface. Therefore, WiFo-D can support downstream inference not only through reconstructed CSI, but also through compact task-transferable quantized representations.

\subsection{Ablation and Scaling Analysis}

We conduct ablation and scaling studies to examine the key
design choices of WiFo-D and its empirical behavior with
increasing pre-training data and model capacity. Both studies
use the 3-bit ID NMSE as the evaluation metric, corresponding
to the most compact discrete representation and the most
stringent feedback setting.

\subsubsection{Component and Training Ablations}

The ablation study evaluates the decoder initialization,
architectural components, and joint multi-codebook training
strategy. Since $N_{\rm c}=8$ corresponds to the 3-bit
feedback setting, the degradation $\Delta$ in
Table~\ref{tab:ablation_3bit} is measured relative to the
standard WiFo-D model under this setting.

The standard model uses Fourier features as the initial decoder
feature $\mathbf{F}_0$, quantized-feature modulation,
attention blocks, and joint multi-codebook training. Although
$\mathbf{F}_0$ is not an independent module, it defines the
reconstruction basis of the decoder. Fourier features
outperform transformed quantized, constant, and positional
initializations, consistent with the coordinate-dependent
oscillations induced by array responses and
frequency-selective propagation.

\begin{table}[!tbp]
\centering
\caption{Ablation results under 3-bit CSI feedback.}
\label{tab:ablation_3bit}
\small
\renewcommand{\arraystretch}{1.2}
\setlength{\tabcolsep}{7.2pt}
\begin{tabular}{@{}lcc@{}}
\noalign{\hrule height 1.0pt}
\textbf{Variant} & \textbf{3-bit ID NMSE (dB)$\downarrow$} & \textbf{$\Delta$ (dB)$\downarrow$} \\
\noalign{\hrule height 0.4pt}
Standard WiFo-D & \textbf{-15.713} & -- \\
\noalign{\hrule height 0.4pt}
\multicolumn{3}{@{}l}{\textit{Decoder initialization}} \\
$\mathbf{F}_0=$ trans. $\mathbf{Z}_{\rm q}$ & -15.424 & +0.289 \\
$\mathbf{F}_0=$ constant & -15.551 & +0.162 \\
$\mathbf{F}_0=$ positional & -15.586 & +0.127 \\
\noalign{\hrule height 0.4pt}
\multicolumn{3}{@{}l}{\textit{Architecture}} \\
w/o attention & -15.057 & +0.656 \\
w/o modulation & -12.852 & +2.861 \\
\noalign{\hrule height 0.4pt}
\multicolumn{3}{@{}l}{\textit{Training strategy}} \\
single 3-bit codebook & -14.863 & +0.850 \\
\noalign{\hrule height 1.0pt}
\end{tabular}
\end{table}

Removing attention degrades the NMSE by 0.656 dB,
confirming the benefit of global antenna-frequency dependency
modeling. For the w/o modulation variant, simply retaining
Fourier features would remove all sample-dependent CSI
information. We therefore feed $\mathbf{Z}_{\rm q}$ directly
to the decoder, concatenate it with the Fourier coordinates,
and remove the modulation generator, thereby comparing
modulation with conventional direct-input decoding. The
resulting 2.861 dB degradation confirms the importance of
modulation for injecting discrete CSI structural priors.

Finally, training only the 3-bit codebook and removing
alternating multi-codebook training degrades the NMSE by
0.850 dB. Joint training therefore not only supports multiple
feedback rates but also improves the low-bit codebook through
the shared encoder and decoder.

\subsubsection{Data and Model Scaling}

\begin{figure}[!tbp]
\centering
\hspace*{-0.035\linewidth}%
\includegraphics[width=0.96\linewidth]{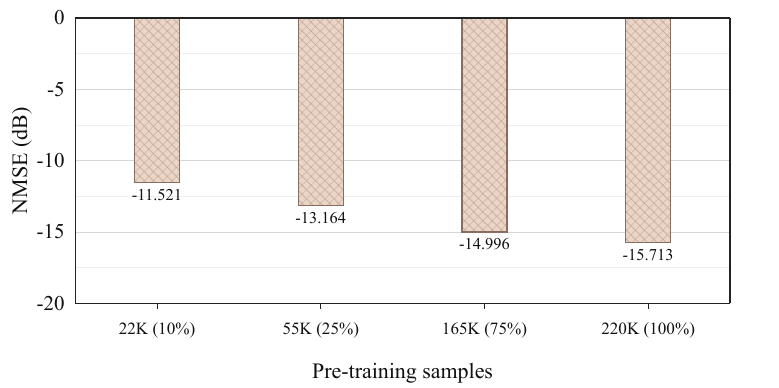}
\caption{Data scaling performance of WiFo-D under 3-bit ID CSI feedback.}
\label{fig:data_scaling}
\end{figure}

\begin{figure}[!tbp]
\centering
\hspace*{-0.035\linewidth}%
\includegraphics[width=0.96\linewidth]
{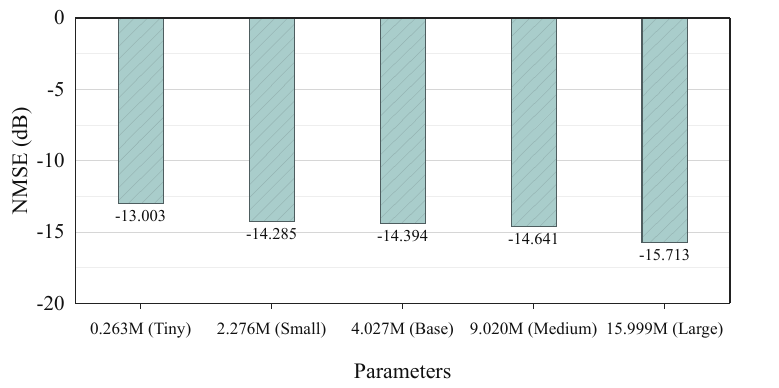}
\caption{Model scaling performance of WiFo-D under 3-bit ID CSI feedback.}
\label{fig:model_scaling}
\end{figure}

We further evaluate whether WiFo-D benefits from increasing
the pre-training data scale and model capacity.
Figs.~\ref{fig:data_scaling} and~\ref{fig:model_scaling}
report the measured 3-bit ID NMSE in dB under different data
and model scales, respectively.

For data scaling, we keep the model architecture unchanged
and vary the proportion of the pre-training datasets. As shown
in Fig.~\ref{fig:data_scaling}, the 3-bit ID NMSE consistently
improves as the number of pre-training samples increases.
Specifically, increasing the data scale from 10\% to 100\%,
corresponding to 22K and 220K pre-training samples,
respectively, improves the NMSE from -11.521 dB to
-15.713 dB, yielding a gain of 4.192 dB. This result
indicates that larger CSI corpora expose the codebooks to
richer spatial-frequency channel patterns and consequently
improve the learned discrete representation.

\begin{table*}[!tbp]
\centering
\caption{Complexity and performance under 3-bit CSI feedback.}
\label{tab:complexity_comparison}
\small
\renewcommand{\arraystretch}{1.2}
\setlength{\tabcolsep}{4.8pt}
\begin{tabular}{l l c c c c c}
\noalign{\hrule height 1.0pt}
\textbf{Model type} &
\textbf{Method} &
\textbf{Params (M)} &
\textbf{FLOPs (G)} &
\textbf{Mean latency (ms)} &
\textbf{Storage (MB)} &
\textbf{3-bit ID NMSE (dB)} \\
\noalign{\hrule height 0.4pt}
\multirow{2}{*}{Task-specific AI model}
& TransNet
& 1.517
& 0.040
& 1.501
& 5.79
& -10.414 \\
& SG-VQ
& 4.744
& 0.307
& 1.190
& 18.10
& -8.625 \\
\noalign{\hrule height 0.4pt}
\multirow{3}{*}{Wireless foundation model}
& WiFo-CF
& 14.183
& 0.728
& 41.279
& 54.10
& -9.391 \\
& WiFo-D-Tiny
& \textbf{0.263}
& 0.057
& 5.873
& \textbf{1.00}
& -13.003 \\
& WiFo-D-Large
& 15.999
& 3.412
& 6.532
& 61.03
& \textbf{-15.713} \\
\noalign{\hrule height 1.0pt}
\end{tabular}
\end{table*}


For model scaling, we construct five WiFo-D variants by
changing the base channel width while keeping the discrete
bottleneck unchanged, including the latent downsampling
factor, codeword dimension, and codebook sizes. The base
channel widths of the Tiny, Small, Base, Medium, and Large
variants are 16, 48, 64, 96, and 128, respectively. The model
used in the main experiments corresponds to the Large
variant. As shown in Fig.~\ref{fig:model_scaling}, increasing
the model size consistently improves the 3-bit ID
reconstruction performance. As the number of parameters
increases from 0.263M to 15.999M, the NMSE improves from
-13.003 dB to -15.713 dB, yielding a gain of 2.710 dB.
Since all variants share the same 3-bit feedback bottleneck,
this gain results from stronger representation learning and
reconstruction capability rather than increased feedback
overhead.

Overall, increasing either the pre-training data scale or model
capacity consistently improves the learned discrete CSI
representation without changing the feedback bottleneck.
These empirical results support the scalability of WiFo-D
with respect to both data and model scale.

\subsection{Complexity and Deployment Analysis}

Complexity is evaluated under 3-bit feedback with an input of
size $1\times2\times32\times32$. FLOPs and latency cover the
complete feedback-reconstruction path, excluding data loading
and metric computation. Latency is measured in FP32 on an
NVIDIA GeForce RTX 5090 using 100 warm-up and 1000 repeated
runs, and storage is reported in FP32. For comparison with
WiFo-D-Large, we use the similarly sized 14.183M-parameter
WiFo-CF configuration.

As shown in Table~\ref{tab:complexity_comparison}, TransNet
and SG-VQ are trained separately for individual domains,
whereas WiFo-CF and WiFo-D share one model across
heterogeneous domains. Their per-model complexity should
therefore be interpreted together with their deployment scope.

WiFo-D-Tiny uses only 0.263M parameters and 1.00 MB storage
while achieving -13.003 dB. It improves over WiFo-CF,
TransNet, and SG-VQ by 3.612 dB, 2.589 dB, and 4.378 dB,
respectively, showing that WiFo-D benefits from its discrete
representation and modulation-based prior rather than merely
from model scale. WiFo-D-Large achieves the best NMSE of
-15.713 dB. Compared with the similarly sized WiFo-CF, it
improves NMSE by 6.322 dB and reduces latency from
41.279 ms to 6.532 ms, despite higher FLOPs and modestly
larger parameter and storage costs
(15.999M versus 14.183M and 61.03 MB versus 54.10 MB).
It still maintains single-digit millisecond latency. Thus,
WiFo-D-Tiny and WiFo-D-Large provide compactness- and
accuracy-oriented operating points, respectively.

WiFo-D further reduces multi-configuration deployment costs
by sharing its encoder, decoder, projections, and
modulation-based backbone across scenarios, antenna/subcarrier
configurations, and feedback rates. Rate adaptation requires
only selecting a codebook. Supporting the six 3-bit to 8-bit
rates with independent WiFo-D-Large backbones would require
approximately $6\times15.999$M parameters, whereas WiFo-D
uses one 15.999M-parameter backbone with negligible
additional codebook storage. It therefore provides both a
favorable complexity--performance tradeoff and a unified
deployment solution across heterogeneous wireless settings.

\section{Conclusion}

This paper proposed WiFo-D, a wireless foundation model that
learns multi-granularity discrete CSI representations
through multi-codebook vector quantization and modulation-based
decoding. The learned token space provides a unified interface for
multi-rate CSI feedback, CSI reconstruction instantiated here by
frequency-domain prediction, and downstream task adaptation. Experiments showed that WiFo-D
achieves strong low-bit ID and OOD feedback performance and
supports 3-bit to 8-bit operation with one shared backbone. It
also improves sample efficiency, noise robustness, and
representative cross-domain transfer in frequency-domain
prediction, while its quantized features outperform
task-specific baselines in beam prediction, wireless
localization, and LoS/NLoS identification. Ablation, scaling, and complexity analyses further validate the
key designs, scalability, and deployment efficiency of WiFo-D.

\end{document}